# Wind Driven Semiconductor Electricity Generator With High Direct Current Output Based On a Dynamic Schottky Junction

*Xutao Yu[1#], Haonan Zheng[1#], Yanghua Lu[1], Jiaqi Si[1], Runjiang Shen[1], Yanfei Yan[1], Zhenzhen Hao[1], Shisheng Lin[1,2]\**

[1]College of Microelectronics, College of Information Science and Electronic Engineering, Zhejiang University, Hangzhou, 310027, China

[2]State Key Laboratory of Modern Optical Instrumentation, Zhejiang University, Hangzhou, 310027, China

[#]These authors contributed equally to this work.

*Email: shishenglin@zju.edu.cn

## Abstract

As the fast development of internet of things (IoTs), distributed sensors have been frequently used and the small and portable power sources are highly demanded. However, the present portable power source such as lithium battery has low capacity and need to be replaced or recharged frequently. A portable power source which can continuously generate electrical power in situ will be an idea solution. Herein, we demonstrate a wind driven semiconductor electricity generator based on a dynamic Schottky junction, which can output a continuous direct current with an average value of 4.4 mA (the maximum value of 8.4 mA) over 360 seconds. Compared with the previous metal/semiconductor generator, the output current is one thousand times higher. Furthermore, this wind driven generator has been explored to function as a turn counter due to its stable output and also to drive a graphene ultraviolet photodetector, which shows a responsivity of 35.8 A/W under the 365 nm ultraviolet light. Our research provides a feasible method to achieve wind power generation and power supply for distributed sensors in the future.

**Main**

With the development of internet of things (IoTs),[1-3] more and more sensors emerge and distribute widely, such as photodetectors,[4-6] temperature sensors and vibration sensors,[7-10] serving in the era of big data. While the rapid development of these scattered power-consuming sensors,[11,12] a smart and in-time energy supply is highly demanded for smooth transmission of big data. Strikingly, energy storage has attracted more attentions than the small and convenient electric generators.[13,14] Nevertheless, lithium battery needs to be frequently recharged or replaced, which is very costly and time-consuming.[15,16] The best way of solving this problem is to create a small and convenient generator,[14,17,18] which for instance make uses of the energy of low-frequency gentle breeze. It is hard to create a small and efficient generator, as clean energy sources like wind energy, tidal energy and geothermal energy can only be transformed into electricity by conventional large size equipment. For example, traditional wind turbines are very complex with a large amount of blades, transmission shaft, gear case, generator unit.[19,20] What's more troublesome is that conventional wind generator unit makes use of heavy coil rotation to cut magnetic inductance line, and produces alternating current.[21-24] In order to store energy and consume energy, generator also needs external rectifying circuit to change it into direct current.[25-29] For powering the distributed sensors in IoTs and making the wind generator more convenient, it is an urgent need to invent a novel type of generator using a simpler device.

Herein, for the first time, we demonstrate a novel wind driven semiconductor electricity generator with ultrahigh direct current output, which is based on a dynamic metal/semiconductor van der Waals Schottky junction only. Our generator has an extremely high direct-current output of 8.4 mA for a contact area of 0.45 cm$^2$ between the metal and semiconductor. Furthermore, this generator has been demonstrated to work continuously over 360 seconds with a stabilized output. As a proof of concept, through a special design with polyimide insulating layer attached to the surface of silicon rod, a portable turn counter has been obtained, whose mechanism is in contrast to the common magnetic turn-counting sensors.[30-32] The wind driven semiconductor electricity generator has also been used to drive a graphene photodetector, which exhibits responsivity over 35.8 A/W. This demonstrates a potential way of direct power supply for driving widely distributed sensors in the age of IoTs.

Figure 1a illustrates the three-dimensional structure of a fabricated wind driven semiconductor electricity generator. The generator is composed of a windmill transmission blade, a semiconductor rotor structure and a metal stator structure, which are mainly supported by the metal bracket to form the complete device structure. It is worth noting that the metal and semiconductor contact closely as the insert of Figure 1b. When the copper sheet is attached with the p-type silicon rod in static state, the static J-V curve of the formed Schottky diode can be measured, which is also shown in Figure 1b. The rectified J-V curve, at the bias voltage from -1.0 V to 1.0 V, demonstrates a good static Schottky junction formed between Cu and p-type silicon. At the same time, the J-V curve of dynamic Cu/p-Si Schottky junction at the bias voltage from -1.0 V to 1.0 V has been shown in the Figure 1c. The wind energy collected is converted into kinetic energy, driving semiconductor rotor structure rotating with the blade. Compared with static J-V curve shown in Figure 1b, it is pretty interesting to find that the dynamic process shows the fluctuation of J-V curve (magnified in the inset of Figure 1c). Certainly, the J-V curve of dynamic Cu/p-Si Schottky also exhibits the good rectification characteristic, which means the fine contact surface of the device. Similar to J-V curve of conventional solar cells under light,[33-35] the J-V curve of our generator in the dynamic process also deviates from the origin, which well confirmed the existence of power generation.

The physical mechanism of this generator can be explained by the dynamic establishment and destruction of the depletion layer, which causes the separation of diffused electrons and holes in the built-in electric field. When metals and semiconductors are in a relative static state, the drift-diffusion equation and current density can be described as follows:

$$J_p = J_{p|drf} + J_{p|dif} = ep\mu_p E - eD_p \nabla p \qquad (1)$$

$$J_n = J_{n|drf} + J_{n|dif} = en\mu_n E + eD_n \nabla n \qquad (2)$$

where $J_p$, $J_{p|drf}$ and $J_{p|dif}$ are the hole current density, the hole drift current density and hole diffusion current density, $J_n$, $J_{n|drf}$ and $J_{n|dif}$ are the electron current density, electron drift current density and electron diffusion current density, $\mu_p$ and $\mu_n$ are the hole/electron mobility, $p$ and $n$ are the position-dependent hole density and electron density in semiconductor, $D_p$ and $D_n$ are the hole/electron diffusion coefficient, respectively. *E* is the built-in electric field, *e* is the elementary charge. In the schematic diagram in Figure 1d, the p-type silicon rod and the copper sheet are in close contact and form a dynamic Schottky junction. The band structure of static Cu/p-Si is illustrated in Figure S1(Supporting Information). The work function of copper is about 4.48 eV,[36] and p-Si is about 5.05 eV (the calculation is listed in the Note S1, Supporting Information). So, when the copper contacts with p-Si surface, a built-in electric field establishes between the Cu and p-Si due to the differences of their work functions. At the same time, depletion layer forms at the interface. When the p-type silicon rod driven by the windmill rotates against the copper sheet, the balance of the static Schottky junction between the interface is broken and a dynamic Schottky junction gradually forms. From the equations (1) and (2), the current density of Cu/Si Schottky junction is composed of $J_{drf}$ and $J_{dif}$. We can assume that the contact area is unchanged during the whole rotation process, which means the origin of power generation is the balance of drift-diffusion broken. As the continuous generation and destruction of the depletion layer, the otherwise diffused electrons and holes lose the paths for diffusing and are reflected or rebounded by the built-in electric field (carriers' movement process are shown in Figure 1d), which forms the direct-current output. Therefore, the fluctuation of J-V curve (Figure 1c) during the dynamic process, which represents the output of power during the dynamic process. When metal and silicon rotate relative to each other continuously, the generator outputs direct-electrical current. The most unique feature in our wind driven semiconductor electricity generator is that it can output an ultrahigh direct current. In Figure 1e and Figure 1f, the peak current and voltage output are 7.6 mA and 67.1 mV, respectively.

Up to now, this output current of 7.6 mA is three orders of magnitude higher than the value reported for the generators based on a moving Schottky diode.[25] Due to the use of wind energy instead of the previous manual sliding, the pressure between the metal copper and the silicon rod has a better fit to generate a larger milliampere level direct current. Therefore, the wind driven generator with a structure of Cu/p-Si based on a dynamic Schottky junction has been realized. As for our generator, more noteworthy is that owing to the output direct-current in a milliamp level and beyond the lower limit of the current analog value 4 mA, which can also be used as the long-distance current transmission analog signal.[37]

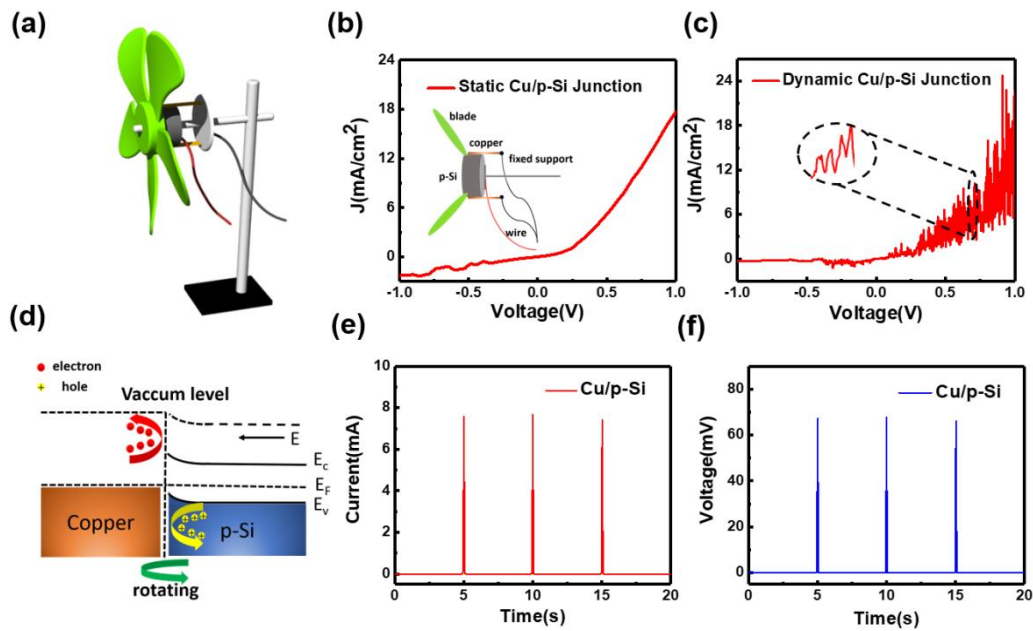

**Figure 1. (a) The three-dimensional model of wind driven semiconductor electricity generator. (b) The J-V curve of the structure of Cu/p-Si generator**

**based on a static Schottky junction at a bias voltage from -1.0 V to 1.0 V. The inset picture is the side view of the rotor structure and stator structure. (c) The rectification characteristic of dynamic Schottky junction from -1.0 V to 1.0 V. (d) The schematic diagram of internal carriers moving process between metal Cu and p-Si based on dynamic Schottky junction. (e) The peak current output and (f) the peak voltage output for dynamic Cu/p-Si wind driven semiconductor electricity generator. All the contact area is 0.4 cm$^2$ between the metal and semiconductor.**

As displayed in Figure 2a, the output voltage and current of generator versus rotating speed has been measured. It can be seen that when the speed gradually increases, the output voltage has a certain positive correlation with the speed, as well as the output current. With the speed increasing from 6 r/s to 11r/s, the voltage increases from 7.2 mV to 70.7 mV and the current increase from 2.3 mA to the maximum value of 8.4 mA. With the rotate speed increasing, it just means that the built-in electric filed is destructed and established rapidly, where the voltage is limited of the barrier height of Cu/p-Si heterojunction simultaneously. Also, the current is limited of by the drift-diffusion equation, where the root of the output current comes from the rebound of diffused carriers. So, when the blades' rotate speed increases further, the output voltage and current of the entire generator saturate gradually.

The power conversion efficiency (PCE) of generator with the structure of Cu/p-Si is also obtained. According to energy conversion relationship, it can be described as follows:

$$I = \frac{1}{2}m(r_1^2 + r_2^2) \qquad (3)$$

$$E_k = \frac{1}{2}I\omega^2 \qquad (4)$$

$$\eta = \frac{P}{E_k} \qquad (5)$$

where $I$ is the rotational inertia of rotor structure, m is the quality of rotor structure, $r_1$ and $r_2$ is the inner diameter and outer diameter respectively, $E_k$ is the kinetic energy of rotor structure, $\omega$ is the rotational angular velocity, $P$ is the output product of output voltage and current, $\eta$ is the PCE of the generator. Based on the equation (3), (4) and (5), the PCE of generator dictates the conversion efficiency of the kinetic energy of the rotor structure into electricity. In the Figure 2b, with the rotate speed varied from 6 r/s to 11 r/s, the PCE continues to increase with a maximum value of 0.56%. In addition to the influence of rotating speed, the contact area between the metal and semiconductor is also a major factor, which influences the electricity output of generator. In order to control the different contact area between the metal and semiconductor, different length of metal sheets have been designed in our generator. In Figure 2c and 2d, the output voltage and current have been obtained with a fixed rotate speed of 11 r/s. The contact area is 0.15 cm$^2$, 0.30 cm$^2$ and 0.45 cm$^2$ respectively. As shown in Figure 2c, the values of output voltage are similar, which exceed over 70.0 mV. The fact that the output voltage depends on the materials' work function difference but not on the contact area, which reflects the mechanism of the generator properly. In Figure 2d, the continuous current output is plotted versus different contact area. The average value of continuous output current increases from 2 mA to 4.7 mA with the increased contact area. From the mechanism of generator, when the contact area of metal and semiconductor increases, the more diffused electrons and holes crossing the depletion layer will be accelerated to bound back to form continuous current output in the dynamic Schottky junction.

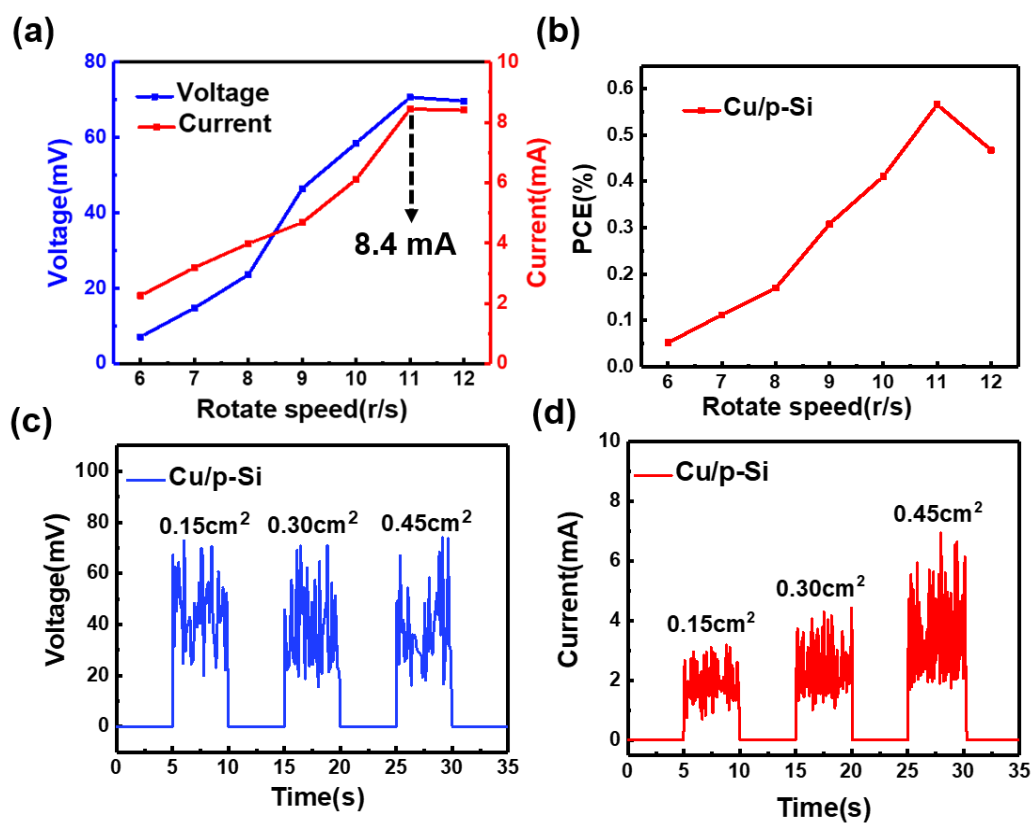

**Figure 2. (a) The correlation of direct voltage and current output on different rotate speed of rotor structure. (b) The dependence of PCE on different rotate speed. (c) The dependence of continuous direct voltage output and (d) the continuous direct current output on contact area at the rotate speed of 11 r/s.**

Furthermore, materials with different work function have been used to test the direct electricity output as shown in the Figure 3a and 3b. The work function of aluminum and n-Si is 4.20 eV and 4.27 eV (the calculation is listed in the Note S1, Supporting Information) respectively. Therefore, the work function of these materials are smaller than that of p-Si, which causes a same direction output electricity due to forming the same direction built-in electric field. A peak of voltage output has been shown among in aluminum, n-Si and copper. For the structure of aluminum/p-Si, it has achieved a direct voltage output with an average value of 0.15 V in Figure 4a. The result can also be repeated as shown in Figure S2 (Supporting Information). The output direct voltage of n-Si is 0.12 V, which is also showed in Figure S3 (Supporting Information). Compared to copper, the differences of work function between and p-Si and aluminum (or n-Si) are nearly 0.2 eV larger. Therefore, the generated voltage of Al or n-Si/p-Si are higher than Cu/p-Si on account of the differences of work function. At the same time, the output current is also measured in Figure 3b. The peak of output current of Cu/p-Si structure is 8.0 mA. The peak of output current of metal aluminum and n-Si is 17.1 μA and 118.0 μA (the comparison of multiple measurements is shown in Figure S4, S5), which are shown in the inset in Figure 4b respectively. Overall, the generator with the structure of Cu and p-Si shows better output characteristics based on a dynamic Schottky junction.

In order to explore practical usage capability, the structure of Cu/p-Si wind driven semiconductor electricity generator is fabricated. The continuous output voltage and current sustained for over 360 seconds has also been shown in Figure 3c. With the rotate speed of 11 r/s and the contact area of 0.45 cm$^2$, the continuous output voltage stabilizes at an average value of 45.0 mV and the continuous output current stabilizes at an average value of 4.4 mA. The stable output direct voltage and current have clearly demonstrated that our generator can continuously and effectively convert low-frequency wind energy into electric energy. Moreover, there is no major damage to the surface of the contact area for the generator with the structure of Cu and p-Si, which shows its potential for practical applications.

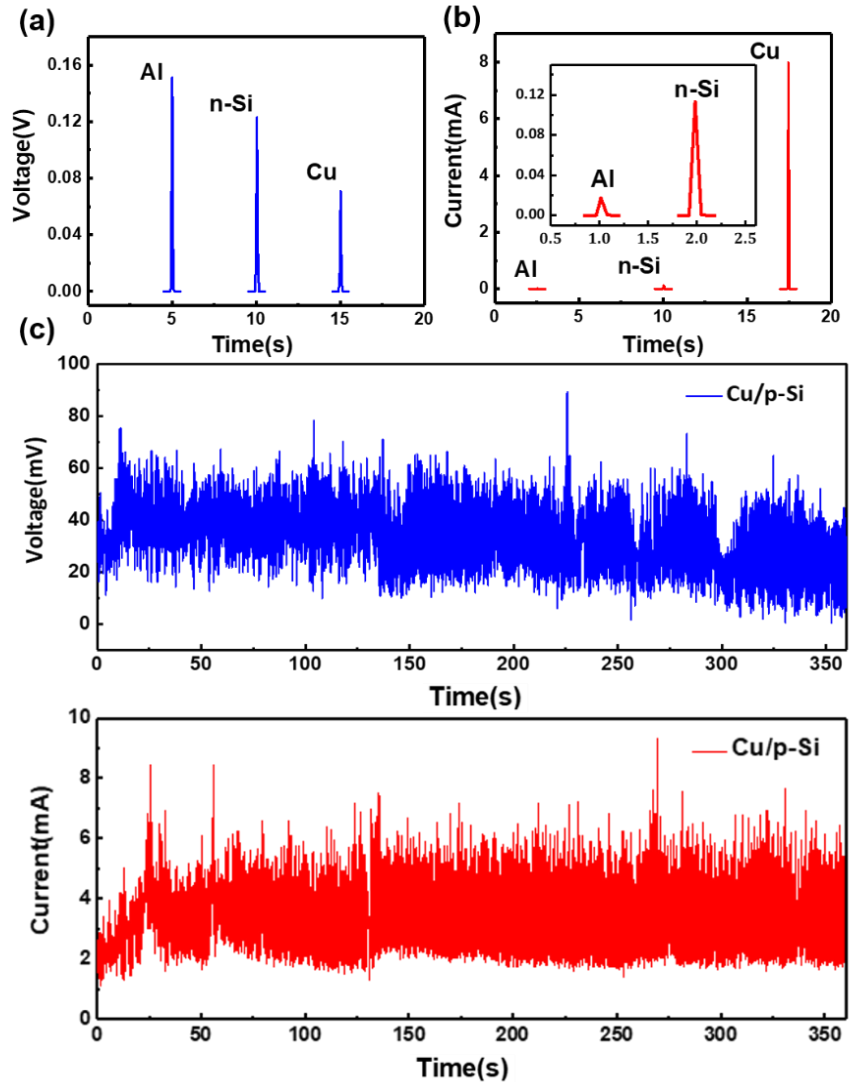

**Figure 3. (a) The voltage output and (b) the current output of generator with different materials. (c) The continuous direct voltage and the direct current output of the Cu/p-Si heterostructure-based wind driven generator over 360 s. All the contact area between rotor structure and stator structure is 0.45 cm$^2$, the rotate speed is 11 r/s.**

In Figure 4, the practical application examples of generator with the structure of Cu/p-Si are present. As shown in Figure 4a, our generator can be used as a turn counter to record the number of turns. In order to achieve the target of counting, a kind of polyimide insulating layer is attached to the surface of silicon, which can separate the metal sheets from the silicon rod surface. Sharp contrast can be achieved between the current peaks and valleys. The number of current peaks is in a good agreement with the number of turns of generator, which means the realization of turn counting by our generator. Furthermore, a graphene/GaN ultraviolet photodetector has been driven by this generator in Figure 4b and 4c. The devices fabrication of graphene/GaN heterostructure is listed in the Note S2 (Supporting Information). The two-dimensional structure diagram and practical device have been shown in Figure S6, S7 (Supporting Information), individually. Under the ultraviolet light of 365 nm, the photodetector powered directly by this generator shows a responsivity of 35.8 A/W and work steadily, which demonstrates the ability of power supply for sensors by our generator.

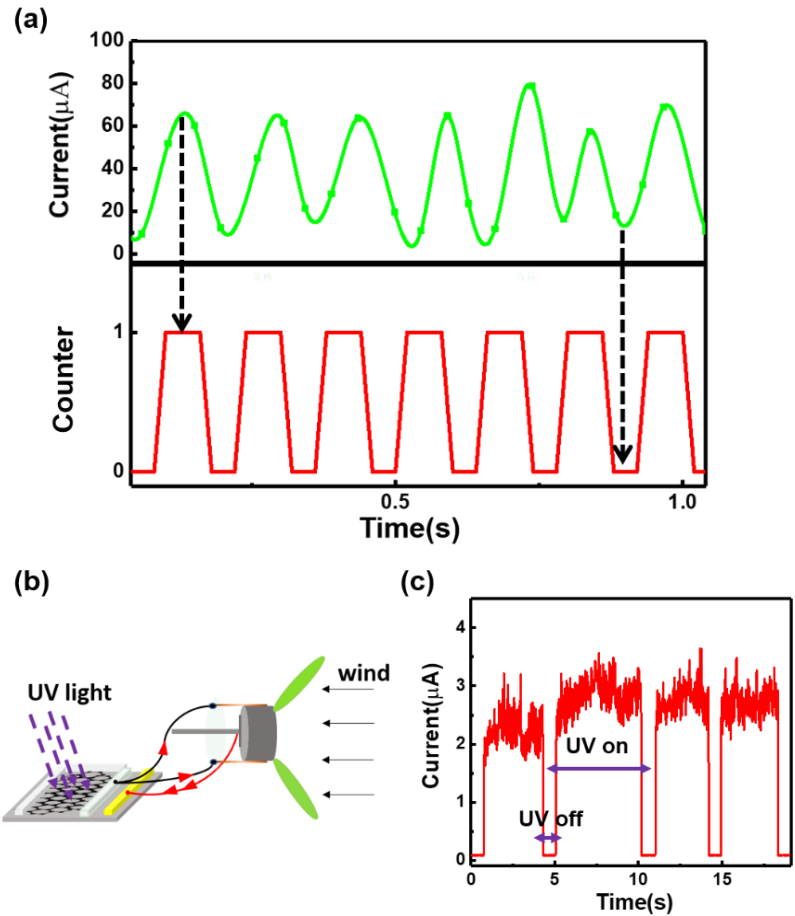

**Figure 4. (a) As a turn counter to count the number of turns of the windmill. (b) The Schematic diagram of generator driving a graphene/GaN ultraviolet photodetector under the 365 nm ultraviolet light. (c) The photoelectric response of graphene/GaN ultraviolet photodetector is driven by our generator under the 365 nm ultraviolet light.**

In summary, a wind driven semiconductor electricity generator based on a dynamic Schottky junction has been demonstrated with a high direct current output. The exotic mechanism is that the dynamic built/disappeared of the built-in electric field rebounded the diffusing carriers to form continuous current output. The key important feature of the generator is that it is only constructed by simple contact of metal and semiconductor, which can reduce the cost of present windmill power generation models through simplifying power generation device and external conversion circuit. With the structure of Cu/p-Si, continuous output voltage/current at an average value of 45.0 mV/4.4 mA over 360 seconds is achieved. The application of generator as a turn counter and power source for graphene ultraviolet sensor is demonstrated successfully, which certifies the great potential of wind power generation supplying for distributed sensors uninterruptedly.

**Experimental section**

***The device fabrication of Cu/p-Si wind driven semiconductor electricity generator:***
The entire experimental setup was made of blade, p-type silicon rod and copper sheet. 55 mm diameter p-type silicon rod was cut into the thickness of 15 mm and washed with acetone, isopropanol, deionized water in sequence to clean surface. After surface treatment, surface silver electrode was prepared and formed good ohmic contact with silicon surface in low contact resistance by annealing in high temperature furnace at 850℃ for 10 minutes. Then, the silicon rod with prepared surface silver electrode was fixed to the blade to form a rotor structure of the windmill. As the stator structure of the windmill, the copper sheet's surface was sequentially removed with ethanol and deionized water in order and complete the device fabrication.

***Physical characterization methods:*** The current-voltage data of were recorded using Keithley 2400 and Agilent B1500A system. The data of current and voltage over time were measured by Keithley 2010 system with a sampling rate of 18 s$^{-1}$. The rotate speed of rotor structure and the contact area between surfaces are acquired by an image processing analysis software Image-Pro Plus (American MEDIA CYBERNETICS), where rotation process of generator is recorded by a video camera.

**Data availability**

The data that support the findings of this study are available from the corresponding author upon reasonable request.

**Acknowledgements**

The authors acknowledge the support from the National Natural Science Foundation of China (No. 51202216, 51502264 and 61774135) and Special Foundation of Young Professor of Zhejiang University (Grant No. 2013QNA5007).

**Author Contributions:** S. S. Lin designed the experiments, analyzed the data and conceived all the works. X. T. Yu and H. N. Zheng carried out the experiments, discussed the results and wrote the paper. Y. H. Lu, J. Q. Si, R. J. Shen, Y. F. Yan and Z. Z. Hao discussed the results and assisted with experiments. All authors contributed to the scientific discussion and the writing of the paper.

**Competing Interests:** The authors declare no competing financial interest. Readers are welcome to comment on the online version of the paper.

**Supplementary Information** is Available in the online website.